
\input phyzzx
\tolerance=5000
\sequentialequations
\overfullrule=0pt
\twelvepoint
\nopubblock
\PHYSREV
\doublespace

\def\a{\alpha}
\def\si{\sigma}
\def\etal{{\it et al.\ }}
\def\dsdr{{d\si\over d\rho}}
\gdef\journal#1, #2, #3, 1#4#5#6{               
    {\sl #1~}{\bf #2}, #3 (1#4#5#6)}            
\def\mpl{\journal Mod. Phys. Lett., }
\def\pl{\journal Phys. Lett., }
\def\prl{\journal Phys. Rev. Lett., }
\def\pr{\journal Phys. Rev., }
\def\prd{\journal Phys. Rev. D, }
\def\prb{\journal Phys. Rev. B, }
\def\ijmp{\journal Int. Jour. Mod. Phys., }

\line{\hfill UdeM-LPN-TH-130}
\line{\hfill hepth/9303078}
\titlepage
\title{Parity violation and the mean field
approximation for the anyon gas}
\author{Didier Caenepeel\footnote{\hbox{$^1$}}{e-mail:
CAENEPEEL@LPS.UMONTREAL.CA}
and Richard MacKenzie\footnote{\hbox{$^2$}}{e-mail:
RBMACK@LPS.UMONTREAL.CA}}
\vskip.2cm
\centerline{{\sl Laboratoire de Physique Nucl\'eaire}}
\centerline{{\sl Universit\'e de Montr\'eal}}
\centerline{{\sl Montr\'eal, Qu\'e, H3C 3J7}}
\centerline{{\sl Canada}}

\abstract{
We examine an approach to justifying the mean field approximation
for the anyon gas, using the scattering of anyons. Parity violation permits
a nonzero average scattering angle, from
which one can extract a mean radius of
curvature for anyons. If this is larger than the interparticle separation,
one expects that the graininess of the statistical magnetic field is
unimportant, and that the mean field approximation is good. We argue that
a non-conventional interaction between anyons is crucial, in which case the
criterion for validity of the approximation is identical to the one deduced
using a self-consistency argument.
}

\endpage

\REF\leimyr{J.M. Leinaas and J. Myrheim,
\journal{Nuo. Cim.}, B37, 1, 1977.}

\REF\wilczekthree{F. Wilczek, \prl 49, 957, 1982.}

\REF\laughlin{R.B. Laughlin, \prl 60, 2677, 1988.}

\REF\fethanlau{A.L. Fetter,
C.B. Hanna and R.B. Laughlin, \prb 39, 9679, 1989.}

\REF\chewilwithal{Y-H. Chen, F. Wilczek,
E. Witten and B. Halperin, \ijmp B3, 1001, 1989.}

\REF\wenzee{X.G. Wen and A.Zee, \prb 41, 240, 1990.}

\REF\bosonbased{H. Mori, \prb 42, 184, 1990;
C. Trugenberger, \prd 45, 3807, 1992;
N. Weiss, \mpl A7, 2627, 1992;
D. Sen and R. Chitra, preprint IISC-CTS-91-1;
D. Boyanovsky and D. Jasnow, \journal Physica, A177, 537, 1991;
D.-H. Lee and M.P.A. Fisher, \prl 63, 903, 1989;
C. Trugenberger, \pl 288B, 121, 1992. For a recent review and references of
both fermion-based and boson-based anyons, see Y. Hosotani, UMN-TH-1106
(1992).}

\REF\zee{A. Zee, ``Semionics: A Theory of High Temperature
Superconductivity'', in {\sl High Temperature Superconductivity\/}, Eds.
K.~Bedell, D.~Pines and J.R.~Schrieffer (Addison-Wesley, 1990).}

\REF\mccmac{J. McCabe and R. MacKenzie, Universit\'e de Montr\'eal preprint
UdeM-LPN-TH-102.}

\REF\refs{G.W. Semenoff and N. Weiss, \pl 250B, 117, 1990; Z.F. Ezawa and
A. Iwazaki, preprint TU-406 (1992); J.L. Goity and J. Soto, preprint
UB-ECM-PF-92-19 (1992).}

\REF\marwil{J. March-Russell and F. Wilczek, \prl 61, 2066, 1988.}

\REF\ahaboh{Y. Aharonov and D. Bohm, \pr 115, 485, 1959.}

\REF\hagen{C.R. Hagen, \prd 41, 2015, 1990.}

\REF\suzsribhalaw{A. Suzuki, M.K. Srivastava, R.K. Bhaduri
and J. Law, \prb 44, 10731, 1991.}

It is common knowledge that the statistics of particles in 2+1 dimensions
can be altered by adding an appropriate ``fictitious'' electric charge and
magnetic flux to each particle.
This gives the particles a so-called
``statistical interaction'' which is, in fact, not a real interaction; it
involves no force. The only effect is in the quantum mechanical phase
between processes which differ only in the windings of the particles around
each other [\leimyr].
Since the statistics of particles is essentially embodied in
such phases (particle exchanges have phase $\pm1$ according to whether the
particles are bosons or fermions), this statistical interaction is aptly
named. Attaching charge $q$ and flux $\Phi$ to the particles results in an
additional contribution per particle exchange of $\exp i q\Phi/2$. Thus, if
$q\Phi$ is an even or odd multiple of $2\pi$ the statistical interaction
introduces no phase or a phase $-1$ per interchange. In the former case it
is trivial; in the latter case (corresponding to one unit of flux)
the statistics of the underlying particles are
changed from Fermi to Bose and vice versa. However for arbirtrary values of
$q\Phi$ the resulting statistics is neither Fermi nor Bose and we have {\it
any\/}ons [\wilczekthree].

There has been much interest recently in the many-anyon system, the anyon
gas.  The analysis is complicated, because even if the
anyons have no conventional interaction, the statistical interaction must
nonetheless be treated as such; in fact, one finds that three-body
interactions arise.

Most discussions of the anyon gas describe the particles as fermions with
statistical interaction, so-called fermion-based
anyons [\laughlin,\fethanlau,\chewilwithal].
The approach is a mean-field
or average-field approximation (MFA), wherein one replaces to lowest order
the statistical magnetic field by a uniform one with the same average
value. Then the difference between the actual and mean magnetic fields is
treated as a perturbation, for instance in the RPA approximation.

Alternatively, one can start with bosons plus statistical interaction,
so-called boson-based anyons [\wenzee,\bosonbased].
Again an MFA is a useful starting point.

One justification for the MFA is a self-consistency argument, given in
Chen, \etal [\chewilwithal]. One assumes
an MFA, after which one has free conventional particles in an
external uniform magnetic field. The classical trajectories are then
circular, and taking a typical particle velocity the radius of curvature
can be calculated. If the number of particles within such a circular orbit
is large, or equivalently if the radius of curvature is larger than the
average interparticle separation, the MFA is deemed to be good.
The result [\chewilwithal,\mccmac] is simply expressed in terms of
the strength of the statistical interaction, which we will call $\alpha$
(so that $\a=0$ --- no statistical interaction ---
corresponds to bosons/fermions for boson/fermion-based
anyons). Then the number of particles inside a typical orbit, $Q$, is
$Q\sim \a^{-2}$ for fermion-based anyons, and $Q\sim\a^{-1}$ for
boson-based anyons. The criterion on $\a$ therefore is
$$
\eqalign{
	\a\ll1\qquad&\hbox{boson-based anyons}\cr
	\a^2\ll1\qquad&\hbox{fermion-based anyons}\cr}
\eqn\howdy
$$
Thus, we see
that the approximation is valid near Bose statistics
for boson-based anyons, and near Fermi statistics for fermion-based anyons;
the difference in powers of $\a$ indicates that one can be slightly further
from conventional statistics for fermion-based anyons before the
approximation breaks down.

Here we would like to explore another possible means of justifying the MFA,
one that is somewhat more sophisticated in that it does not rely on
self-consistency. The basic idea, hinted at in Ref. [\wenzee],
arises due to parity violation, which is a
characteristic of generic anyon models (although models can be constructed
which eliminate parity violation [\refs]). Among other things, parity
violation allows the possibility of an asymmetry in the scattering of
anyons: the scattering cross section is not necessarily an even function of
the scattering angle $\rho$, and in a collision between two anyons, on
average a particle would scatter preferentially to the left, say. (Such a
situation can't happen if parity is conserved, in which case $d\si/d\rho$
must be an even function of the scattering angle $\rho$.) An explicit
calculation of the scattering of anyons with an additional hard-core
repulsion, which nicely illustrates the asymmetry in the scattering, has
been performed by Suzuki, \etal [\suzsribhalaw].

Consider the trajectory of an individual particle in a gas of anyons. This
preferential handedness of the scattering will mean that
the particle will have
a roughly circular orbit; we can ascribe a mean radius of curvature for the
particle, a sort of discretized analog of the mean-field radius of
curvature mentioned above. As in that case, if the number of other
particles within the orbit is large, one can say that the graininess of the
magnetic field (a sum of delta functions at the particle positions) is
relatively unimportant and the MFA is deemed to be good.

We initially
consider free anyons for simplicity, which is essentially pure
Aharonov-Bohm scattering [\ahaboh,\hagen], although we will see that the
addition of an additional conventional
interaction appears to be necessary. From the identical
two-anyon scattering cross-section, we derive an expression for the mean
radius of curvature of a typical particle trajectory, and find conditions
under which the radius is bigger than the average distance between
particles, whence the MFA is declared to be good.

One expects that when the scattering is weak, \ie, when the statistical
parameter is small, the average scattering angle will be small and
consequently the radius of curvature large; hence it seems reasonable that
for weak statistical interaction, the MFA will be seen to be justified.
This was just what the self-consistent argument mentioned above concluded.
Furthermore, one expects that in the domain where both the self-consistent
argument and the argument presented here  indicate that the MFA is valid,
the radii of curvature would agree. This is also found to be true (modulo
the necessity for an additional interaction).

Our starting point is
the result of March-Russel and Wilczek [\marwil] (MRW) for the
scattering of boson-based anyons with an additional, conventional
interaction, which is parameterized by a phase shift, $\delta$. Their
result is, for the center-of-mass scattering of two identical anyons,
$$
\dsdr=R+I,
\eqn\un
$$
where
$$
\eqalign{
	R&={4(1-\cos\delta)\over \pi k}
	+{2\sin^2\a\pi\over \pi k\sin^2\rho},\cr
	I&=-{8\sin\a\pi\sin(\delta/2)\over\pi k\sin\rho}
	\sin(|\a\pi|-\delta/2+sgn(\a\pi)\rho).\cr}
$$
Here $\a$ is the strength of the statistical parameter (defined so that, for
boson-based anyons, even/odd integers correspond to bosons/fermions;
MRW's $\theta_v$ corresponds to our $\a\pi$),
$k$ is the particle
momentum, and $\rho$ is the scattering angle.

One peculiarity of anyon scattering, as remarked by MRW, is that  if we
consider the case of free anyons, $\delta=0$, the interference term $I$
goes away and we get a conventional-looking scattering:
$$
\left(\dsdr\right)_{\delta=0} ={2\sin^2\a\pi\over\pi k\sin^2\rho}.
$$
In particular, {\sl there is no parity violation in the scattering of
free anyons.} This means that there will be no preferential handedness to
the anyon trajectories: they travel in straight lines, on average. This is
in stark contrast with their circular trajectories after the MFA.
Naively following the line of reasoning outlined above, we
would conclude that the radius of curvature of such a trajectory is
infinite, and thus that an infinity of particles lie `inside' the
trajectory, so that the MFA is, in a sense,
infinitely justified. However it is,
initially, at least, rather
unsatisfying that the trajectories before and after the MFA
is made have little in common. For the most part in what
follows, we will assume an interaction among the particles, commenting in
the end on the free case.

Thus, let us calculate the mean radius of curvature for a typical
particle trajectory from \un. An ingredient in the calculation is certainly
the following quantity:
$$
X\equiv\int d\rho\,\rho\dsdr.
$$
The mean scattering angle is simply
this, normalized by the cross section:
$$
\bar\rho={\int d\rho\,\rho\dsdr\over\int d\rho\,\dsdr}
={X\over\si}.
$$
We need to supplement this with a mean length between scatterings. This is
$$
L={1\over s\si},
$$
where $s$ is the scattering source density, \ie, the density of anyons in
the gas.

On average, the particle travels a distance $L$ before scattering, and
its scattering angle is $\bar\rho$. If the angle is small, the radius of
curvature is
$$
\bar R={L\over|\bar\rho|}={1\over s|X|}.
\eqn\deux
$$
The number of particles contained within this roughly-circular orbit will
then be
$$
Q\sim {\bar R}^2 s={1\over s X^2};
$$
the MFA is justified if $Q\gg1$, \ie, if
$$
sX^2\ll1.
\eqn\trois
$$

Now we are equipped to calculate $X$, using \un. Since $R$ is
an even function of $\rho$, its contribution to $X$ integrates to
zero. We are left with a contribution from $I$, which gives, eventually,
$$
X=-{8\log 2\over k}\sin\a\pi\sin{\delta\over2}
\sin\left(|\a\pi|-{\delta\over2}\right).
\eqn\quatre
$$
Here we see clearly that if $\a$ is an integer
(corresponding to fermions or
bosons), $X$ is zero, as expected (due to parity). Furthermore, if
$\delta=0$ (corresponding to free anyons), $X=0$, which requires some
interpretation, as
will be discussed below.

The condition for validity of the MFA, \trois, becomes
$$
s\left({8\log 2\over k}\sin\a\pi\sin{\delta\over2}
\sin\left(|\a\pi|-{\delta\over2}\right)\right)^2\ll1.
$$
This unweildy expression can be simplified by
supposing that we have a `generic' (in particular, not small) interaction.
Then we can take $\sin\delta/2$ and
$\sin(\a\pi-\delta/2)$ of order one, and \quatre\ becomes, ignoring numerical
factors,
$$
s\left({\sin\a\pi\over k}\right)^2\ll1.
\eqn\cinq
$$
compared with the self-consistent result for boson-based anyons, $\a\ll1$.

We need an estimate for $k$, the minimum momentum allowed (since we are
interested in the ground state).
This can be obtained in the following way. The
mean radius of curvature gives us a length scale for the particle's trajectory
if we think classically; quantum mechanically, we can say that the wave
function is localized in a region of this order. The uncertainty principle
then gives us a minimum momentum of $k\sim 1/\bar R\sim s\sin\a\pi/k$, using
\deux\ and \quatre, so $k^2\sim s\sin\a\pi$. In
\cinq, this gives
$$
\sin\a\pi\ll1.
\eqn\six
$$
 From this, we must have either $\a$ near zero, corresponding to
a weak statistical
interaction and anyons near bosons, or $\a$ near $1$, corresponding to
anyons
near fermions. (We recall that the above calculation is performed for the
case of boson-based anyons.) In the former case, the criterion reduces to
$$
\a\ll1,
\eqn\sept
$$
which is the same condition as obtained using
the less-sophisticated self-consistent method ({\it cf.} \howdy).

Intuitively, our result
seems very plausible: when $\a$ satisfies \six, the anyons are very near
either bosons or fermions, in which case we expect that parity violation
will be minimal and the average scattering angle will be small. Thus the
mean radius of curvature of the anyon trajectory will be large and the
mean-field approximation is consistent.

It is perhaps noteworthy that previous discussions of the validity of the
MFA have concluded that the approximation is valid for anyons of statistics
near that of the base particle. Here, we find that, for boson-based anyons
(with `generic' interaction),
either statistics near Bose or Fermi yield a justified MFA. This is very
satisfying, since one expects that the physics of anyons should be
independent of whether the underlying particles are bosons or fermions.

The case of free anyons is rather peculiar. Taken at face value, our result
seems to indicate that the MFA is indeed a very bad approximation for this
case. The scattering of free anyons does not violate parity
[\marwil], {\sl no matter what the statistical paramater}.
Thus, the trajectory of an anyon is apparently straight (on average),
in stark contrast to the situation after the MFA, where the particle
trajectories are circular classically.

However we have perhaps made an approximation which is too crude, at least
for the case of free anyons. Namely, we have considered the motion of one
anyon in the gas to be an uncorrelated sequence of two-particle scattering
processes. This neglects any interference effects between subsequent
scatterings, which intuitively should be present, and which would be
non-negligible at least at high density. It is plausible that interference
terms which arise (if we, schematically, consider the amplitude for several
scatterings and then square to get a scattering probability) would affect
the result significantly. In contrast, if there is an interaction the
relative phases are incoherent and interference terms will average to zero.

A more honest study of this problem would be, for instance, to investigate
the movement of a charged particle in the presence of a lattice of flux
tubes. This appears to be quite difficult.

To summarize, we have considered parity violation in the scattering of
boson-based anyons as a means of providing a criterion for the validity of
the mean field approximation in an anyon gas. Adding a conventional
interaction, we find that the mean radius of the anyon trajectory is large,
and thus the graininess of the anyons is unimportant, for anyons near
either bosons or fermions. For the former case, the criterion actually
turns out to be the same as the less-sophisticated self-consistency
calculation, solidifying our confidence in this result.

We gratefully acknowledge very useful discussions with J. McCabe and F.
Wilczek. This work was supported in part by the Natural Science and
Engineering Research Council of Canada and the Fonds F.C.A.R. du Qu\'ebec.

\refout

\end